\documentclass{ws-procs961x669}            % default, citations in superscript
\begin{document}
\hspace*{8cm}
\hfill CERN-TH-2016-110
\title{Cosmology and Supergravity$^*$}

\author{S. Ferrara${}^{a,b,c}$ }

\address{${}^a$Theoretical Physics Department, CERN,\\
CH-1211 Geneva 23, SWITZERLAND\\
$ {}^b$INFN-Laboratori Nazionali di Frascati \\
 Via Enrico Fermi 40, 00044 Frascati ITALY\\
 ${}^c$Department of Physics and Astronomy, U.C.L.A.,  \\
 Los Angeles, CA 90095-1547 USA\\
 E-mail: Sergio.Ferrara@cern.ch}

\author{A. Kehagias}

\address{Physics Division, NTU Athens,\\
15781 Zografou, Athens, GREECE\\
E-mail: kehagias@central.ntua.gr}

\author{A. Sagnotti}
\address{
Scuola Normale Superiore and INFN \\
Piazza dei Cavalieri 7
56126 Pisa ITALY\\
E-mail: sagnotti@sns.it}

\begin{abstract}
 Abdus Salam was a true master of 20th Century Theoretical Physics.
 Not only was he a pioneer of the Standard Model (for which he shared the Nobel Prize with S.~Glashow and S.~Weinberg), but he also (co)authored many other outstanding contributions to the field of Fundamental Interactions and their unification. In particular, he was a major contributor to the development of supersymmetric theories, where he also coined the word “Supersymmetry” (replacing the earlier ``Supergauges" drawn from String Theory). He also introduced the basic concept of ``Superspace" and the notion of ``Goldstone Fermion"(Goldstino). These concepts proved instrumental for the exploration of the ultraviolet properties and for the study of spontaneously broken phases
of super Yang-Mills theories and Supergravity. They continue to play
a key role in current developments in Early-Universe Cosmology. In this contribution we review models of inflation based on Supergravity with spontaneously broken local supersymmetry, with emphasis on the role of nilpotent superfields to describe a de Sitter phase of our Universe.

\end{abstract}


\vskip 2.5cm

\def\thefootnote{\arabic{footnote}}
\setcounter{footnote}{0}
\vfill
\vskip.2in
\noindent
\line(1,0){250}\\
{\footnotesize {$^*$
Contribution to the Proceedings of the ``Memorial Meeting for Nobel Laureate
Prof.~Abdus Salam's 90th Birthday",  Nanyang Technological University, Singapore, 25-28 January 2016.
}}

\bodymatter

\section{Introduction}
Supergravity ~\cite{sugra} combines Supersymmetry with General Relativity (GR). This brings about scalar fields, some of which can play a natural role in the Early Universe.
Nowadays it is well established that inflationary Cosmology is accurately described via the evolution of a single real scalar field, the \emph{inflaton}, in a Friedmann, Lema\^itre, Robertson, Walker (FLRW) geometry ~\cite{cosmo_rev}. A scalar field associated to the Higgs particle was also recently discovered at LHC ~\cite{LHC}, confirming the interpretation of the Standard Model as a spontaneously broken
phase (BEH mechanism) of a non-abelian Yang-Mills theory ~\cite{BEH}.
There is thus some evidence that Nature is inclined to favor, both in Cosmology and in Particle Physics, theories with scalar degrees of freedom, albeit in diverse ranges of energy scales.

Interestingly, there is also a cosmological model where \emph{inflaton} and \emph{Higgs} fields are identified: this is the Higgs inflation model of ~\cite{BS}, which rests on a non-minimal coupling $h^2 \, R$ of the Higgs field  $h$  to gravity.
Another well--known example rests on an $R + R^2$ extension of General Relativity (GR). This is the Starobinsky model of inflation~~\cite{Starobinsky,CM},
which is also conformally equivalent to GR coupled to a scalar field, the \emph{scalaron} ~\cite{Whitt},
with the special scalar potential
\begin{eqnarray}
V=V_0\left(1\ - \ e^{\,-\sqrt{\frac{2}{3}}\,\phi}\right)^2\ , \qquad V_0\sim 10^{-9} ~\mbox{in Planck units \ .}
\end{eqnarray}
These two models (and also a more general class) give identical predictions ~\cite{KDizR} for the slow-roll parameters $\epsilon$ and $\eta$, which are determined by the potential according to
\begin{eqnarray}
\epsilon\ = \ \frac{M_P^2}{2}\left(\frac{V'}{V}\right)^2\ , \qquad \eta\ =\ M_P^2\ \frac{V''}{V} \ .
\end{eqnarray}
The spectral index of scalar perturbations (scalar tilt) and the tensor-to-scalar ratio turn out to be
\begin{eqnarray}
n_s\ =\ 1\ -\ 6\,\epsilon\ +\ 2\,\eta\ \simeq \ 1\ -\ \frac{2}{N} \ , \label{ns} \qquad
r\ =\ 16\,\epsilon\ \simeq \ \frac{12}{N^2} \ , \label{r}
 \end{eqnarray}
where
\begin{eqnarray}
N\ =\ \frac{1}{M_P^2}\int_{\phi_{end}}^\phi \frac{V}{V'}\ d \phi
\end{eqnarray}
is the total number of $e$-folds of inflation.

An interesting modification of the Starobinsky potential, suggested by its embedding in $R + R^2$ Supergravity ~\cite{FKR,FKLP},
involves a deformation parameter
$\alpha$ and reads ~\cite{FKLP,KLR}
\begin{eqnarray}
V_\alpha=V_0\left(1\ -\ e^{\,-\sqrt{\frac{2}{3\alpha}}\,\phi}\right)^2 \ . \label{Valpha}
\end{eqnarray}
It gives the same result of eq.~(\ref{ns}) for $n_s$, but the tensor-to-scalar ratio is now
\begin{eqnarray}
r\ \simeq \ \frac{12\,\alpha}{N^2} \ .
\end{eqnarray}
This family of models provides an interpolation between the Starobinsky model (for $\alpha=1$) and Linde's chaotic inflation model ~\cite{chaotic}, with a quadratic potential (in the limit $\alpha \to \infty$).
The chaotic inflation model leads again to the scalar tilt (\ref{ns}), but now the tensor-to-scalar ratio becomes
\begin{eqnarray}
r\ \simeq \ \frac{8}{N} \ . \label{r_chaotic}
\end{eqnarray}
The recent 2015 data analysis from Planck ~\cite{Planck} and BICEP2 ~\cite{BICEP2} favors  $n_s \approx 0.97$ and  $ r < 0.1$, and thus the Starobinsky model, which lies well within the allowed parameter space due to the additional $1/N$ suppression factor $r$ present in eq.~(\ref{r})
as compared to eq.~(\ref{r_chaotic}).

The form (\ref{Valpha}) for $V_\alpha$ can be further generalized, allowing for an arbitrary, monotonically increasing function $f\left(\tanh \frac{\varphi}{\sqrt{6\alpha}}\right)$,
such that
\begin{eqnarray}
V_\alpha=V_0\ f\left(\tanh \frac{\varphi}{\sqrt{6\alpha}}\right)^2, ~~~\varphi\to \infty: ~~
f\left(\tanh \frac{\varphi}{\sqrt{6\alpha}}\right)\ \to \ 1\ - \ e^{\,-\sqrt{\frac{2}{3\alpha}}\, \phi}\ +\ \ldots \ .
\end{eqnarray}
These modifications led to the concept of $\alpha$-attractors
~\cite{KLR}.
%(Kallosh, Linde, Roest).

This contribution is organized as follows: In section 2 we describe the single-field inflation in Supergravity, in section 3 we discuss inflation and  supersymmetry breaking and in section 4 we present some minimal Supergravity models of inflation.   Nilpotent superfields and sgoldstino-less models are reviewed  in section 5, in section 6  we discuss  higher-curvature Supergravity and its dual standard Supergravity description, in section 7 orthogonal nilpotent superfields are explored  and section 8 contains our conclusions and outlooks.  Finally in appendix A we  briefly review constraint superfields which preserve ${\cal N}=1$ supersymmetry.

\section{Single--Field Inflation in Supergravity}

We can now describe how ${\cal N}=1$  Supergravity can accommodate these ``single--field'' inflationary models, explaining how to embed the \emph{inflaton} $\varphi$ in a general Supergravity theory coupled to matter in an FLRW geometry and the role of its superpartners. Under the assumption that no additional Supersymmetry (${\cal N}\geq 2$) is restored in the Early Universe, the most general  ${\cal N}=1$ extension of GR is obtained by coupling the graviton multiplet  $(2,3/2)$  to a certain number of chiral multiplets $(1/2,0,0)$, whose complex scalar fields are denoted by $z^i, i=1,\ldots,N_s/2$ and to (gauge) vector multiplets $(1,1/2)$, whose vector fields are denoted by
$A_\mu^\Lambda ~(\Lambda=1,\ldots,N_V).$
These multiplets can acquire supersymmetric masses, and in this case the massive vector multiplet becomes $(1,2(1/2),0)$, eating a chiral multiplet in the supersymmetric version of the BEH mechanism.

For Cosmology, the relevant part of the Lagrangian
~\cite{CFGvP,BW}
% (Cremmer, SF, Girardello, Van Proeyen; Bagger, Witten)
is the sector that couples the scalar fields to the Einstein--Hilbert action, described by
\begin{eqnarray}
{\cal L}\ =\ -\ R\ -\ \partial_i\partial_{\overline \j}\, K\,  D_\mu z^i \, D_\nu \bar z ^{\overline \j} \, g^{\mu\nu}\ -\ V(z,\overline z)\ +\ \ldots\ ,
\end{eqnarray}
where $K$ is the K\"ahler potential of the $\sigma$-model scalar geometry and the ``dots" hide fermionic terms and gauge interactions. The scalar covariant derivative is
\begin{eqnarray}
D_\mu z^i\ =\ \partial_\mu z^i \ +\ \delta_\Lambda z^i A_\mu ^\Lambda \ ,
\end{eqnarray}
where the $\delta_\Lambda z^i$ are Killing vectors. This term allows to write massive vector
multiplets  {\it \`{a} la} Stueckelberg.
The scalar potential is
\begin{eqnarray}
V(z^i ,\overline z ^{\overline i})\ =\ e^G\bigg[G_i\,G_{\overline \j}\, (G^{-1})^{i\overline \j}\ -\ 3\bigg]
\ +\ \frac{1}{2}\ (\mathit{Re}f_{\Lambda \Sigma})^{-1}D_\Lambda \, D_\Sigma \ , \label{pot}
\end{eqnarray}
where, in terms of the superpotential $W(z^i)$,
\begin{eqnarray}
G\ =\ K \ +\ \log|W|^2\ , \qquad G_{i\overline j} \ = \ \partial_i\,\partial_{\overline j}\, K\ .
\end{eqnarray}

The first and third non--negative terms in eq.~(\ref{pot}) are usually referred to as ``F" and ``D" term contributions: together with the second, negative term, they encode the option of attaining unbroken Supersymmetry in Anti-de Sitter space.
Alternatively, the potential can be recast in the more compact form
\begin{eqnarray}
V(z^i ,\overline z ^{\overline \i})\ =\ F_i F^{ i}\ +\ D_\Lambda D^\Lambda\ -\ 3 \,|W|^2 \,e^K\ ,
\end{eqnarray}
where
\begin{eqnarray}
F_i\ =\ e^{\frac{K}{2}}\Big{(} W K_{,i}\ +\ W_{,i}\Big{)}\ , \qquad D_\Lambda\ =\ G_{,i}\, \delta_\Lambda z^i\ .
\end{eqnarray}

The D-term potential can endow a vector multiplet with a supersymmetric mass term, and can also give rise to a de Sitter phase, thanks to its non--negative contribution to the potential. Only F-breaking terms can thus give AdS phases. The (field dependent) matrices  $\mathit{Re}f_{\Lambda \Sigma}, ~\mathit{Im}f_{\Lambda \Sigma}$ provide the normalization of the terms quadratic in Yang-Mills curvatures. Their role in Cosmology deserves to be investigated further, since they give direct couplings of the \emph{inflaton} to matter, which are relevant for the epoch of reheating.

\section{Inflation and Supersymmetry Breaking}

In a given phase, unbroken Supersymmetry requires
\begin{eqnarray}
F_i\ =\ D^\Lambda\ =\ 0\ ,
\end{eqnarray}
so that
\begin{eqnarray}
V\ =\ -\ 3\,|W|^2\, e^K\ .
\end{eqnarray}
These are Minkowski or AdS phases depending on whether or not $W$  vanishes. On the other hand, supersymmetry is broken if at least one of the $F_i$ or $D^\Lambda$  does not vanish. In phases with broken Supersymmetry one can have maximally symmetric AdS, dS or Minkowski vacua, so that one can accommodate both the inflationary phase (dS) and the subsequent Particle Physics (Minkowski) phase. However, it is not trivial to construct corresponding models, since the two scales are very different if Supersymmetry is at least partly related to the Hierarchy problem.

In view of the negative term present in the scalar potential (\ref{pot}) it might seem impossible (or at least not natural) to retrieve a de Sitter phase for large values of a scalar field to be identified with the \emph{inflaton}. The supersymmetric versions of the $R+R^2$ (Starobinsky) model show how this puzzle is resolved: either the theory has (with F-terms) a no-scale structure, which makes the potential positive along the inflationary trajectory
~\cite{Cecotti}, or the potential is a pure D-term and is therefore positive ~\cite{CFPS}.
 %(Cecotti, SF, Porrati, Sabharwal) .

These models contain two chiral superfields $(T,S)$ ~\cite{ENO,Kallosh-Linde},
 as in the old minimal version of $R+R^2$ Supergravity ~\cite{Cecotti},
 or one massive vector multiplet ~\cite{FKR,FKLP}, as in the new minimal version, and attain unbroken Supersymmetry in a Minkowski vacuum at the end of inflation.

In the framework of nilpotent superfield inflation ~\cite{ADFS}, some progress was recently made ~\cite{Kallosh-Linde2,DAZ} on the problem of embedding two different supersymmetry breaking scales in the inflationary potential. The multiplet $S$, which does not contain the \emph{inflaton} ($T$ multiplet), is replaced by a nilpotent superfield satisfying
\begin{eqnarray}
S^2\ = \ 0 \ . \label{nilpotent}
\end{eqnarray}
This condition  eliminates the \emph{sgoldstino} scalar from the theory, but its F-component still drives inflation, or at least participates in it.
This mechanism was first applied to the Starobinsky model, replacing the $S$ field by a Volkov-Akulov nilpotent field~~\cite{ADFS}
and then to general F-term induced inflationary models~~\cite{FKL}.
Although the examples are so far restricted to the ${\cal N}=1 \to {\cal N}=0$ breaking in four--dimensional supergravity, these types of construction are potentially very instructive for String Theory, where one readily looses control of the vacuum in the presence of broken supersymmetry~~\cite{resum}. Orientifold vacua~~\cite{orientifolds} provide a natural and interesting entry point into this intricate dynamics, via the phenomenon of ``brane SUSY breaking”~~\cite{bsb}. This rests on non--BPS combinations of branes and orientifolds that are individually BPS, and its simplest ten--dimensional setting was related to non--linear supersymmetry in~~\cite{dm}. Recent work, starting from ref.~~\cite{FKL}, linked it more clearly to the superHiggs effect in Supergravity~~\cite{CFGvP}, and also to the KKLT scenario of~~\cite{KKLT}.
Let us conclude this section, however, by recalling that a first attempt to make use of the nilpotent Volkov-Akulov multiplet in Cosmology, identifying the \emph{inflaton} with the \emph{sgoldstino}, was made in ref.~~\cite{A-GGJ}.
%Alvarez-Gaume,Gomez,Jemmies as quoted in ADFS

\section{Minimal Models for Inflation and Supergravity}

This class includes models where the inflaton is identified with the sgoldstino and only one chiral multiplet $T$ is used. However, the $f(R)$ Supergravity models
% (Ketov)
~\cite{KS1}  yield potentials that either have no plateau or, when they do, lead to
AdS rather than to dS phases~~\cite{FKvP,FKP}.
% (SF,Kallosh, Van Proeyen; SF, Kehagias, Porrati).
This also reflects a no-go theorem ~~\cite{avatars}
%(Ellis, Nanopoulos, Olive).

A way out of this situation was recently found with ``$\alpha$-scale
Supergravity"~~\cite{RS}:
%(Roest, Scalisi)
adding two superpotentials $W_+ + W_-$
which separately give a flat potential along the inflaton ($\mathit{Re}T$) direction can result in a de Sitter plateau for large $\mathit{Re}T$. The problem with these models is that the inflaton trajectory is unstable in the ImT direction, but only for small inflaton field: modifications to the superpotential are advocated to generate a satisfactory inflationary potential. For single-field models and related problems, see also~~\cite{KTer}.
%( Ketov, Terada).
$R+R^2$ Supergravity, D-term inflation~~\cite{FKLP, FFS},
%(SF, Kallosh, Linde, Porrati; SF, Fré, Sorin),
$\alpha$-attractor scenarios
~\cite{CKLR},
%(Kallosh, Linde, Roest, Carrasco),
 no-scale inflationary models
~~\cite{ENO},
 % (Ellis, Nanopoulos, Olive)
 and $\alpha$-scale models
~~\cite{RS}
 %(Roest, Scalisi)
 have a nice $SU(1,1)/U(1)$ hyperbolic geometry for the inflaton superfield, with
\begin{eqnarray}
R_\alpha\ =\ -\ \frac{2}{3\alpha}\ , \qquad n_s\ \simeq \ 1\ -\ \frac{2}{N}\ , \qquad r\ \simeq \ \frac{12\,\alpha}{N^2}\ ,
\end{eqnarray}
where $R_\alpha$ is the curvature of the scalar manifold.
\subsection{D-term Inflation}

An appealing and economical class of models allows to describe any potential of a single scalar field which is the square of a real function
%(SF, Kallosh, Linde, Porrati):
~\cite{FKLP}:
\begin{eqnarray}
V(\varphi)\ = \ \frac{g^2}{2} \ P^2(\varphi)\ .
\end{eqnarray}
These are the D-term models, which describe the self-interactions of a massive vector multiplet whose scalar component is the inflaton. Up to an integration constant (the Fayet-Iliopoulos term), the potential is fixed by the geometry, since the K\"ahler metric is
\begin{eqnarray}
ds^2\ =\ d\varphi^2\ +\ \Big{(}P'(\varphi)\Big{)}^2 da^2\ .
\end{eqnarray}
After gauge fixing, the field $a$ is absorbed by the vector, via $da+gA$, giving rise to a mass term
$ \frac{g^2}{2}  \Big{(}P'(\varphi)\Big{)}^2 A_\mu^2$ (BEH mechanism).
In particular, the Starobinsky model corresponds to
\begin{eqnarray}
P(\varphi)\ =\ 1\ -\ e^{\,-\sqrt{\frac{2}{3}}\,\varphi}\ ,
\end{eqnarray}
but in all these examples there is no superpotential and only a de Sitter plateau is possible. At the end of inflation $\varphi=0$, $D=0$ and Supersymmetry is recovered in Minkowski space, since $V=0.$

 \subsection{$R+R^2$ Supergravity}

There are two distinct classes of models, depending on the choice of auxiliary fields: old and new minimal models. The off-shell degrees of freedom contain the $6 (= 10-4 {\rm diff})$ degrees of freedom of the  graviton $g_{\mu\nu}$ and the $12 (=16-4 {\rm diff})$ degrees of freedom of the gravitino $\psi_\mu$. The $n_B=n_F$ off-shell condition requires six more bosons. There are two choices for the latter, which reflect the two minimal supegravity multiplets of the ${\cal N}=1$ theory:
\begin{itemize}
\item old minimal: ~~~$A_\mu, ~S,~P ~$(6 DOF's)
\item new minimal: ~~~$A_\mu, B_{\mu\nu}~$ (6 DOF's due to gauge inv. $\delta B_{\mu\nu}=\partial_\mu b_\nu-\partial_\nu b_\mu$) \ .
\end{itemize}
These $12_B+12_F$ degrees of freedom must fill massive multiplets like
\begin{eqnarray}
 {\rm Weyl}^2:~(2,2(3/2),1), ~~~R_{\rm old}^2:~2(1/2,0,0), ~~~R_{\rm new}^2: ~ (1,2(1/2),0).
 \end{eqnarray}

After superconformal manipulations, these two theories can be turned into standard Supergravity coupled to matter. The new minimal gives D-term inflation as described before, while the old minimal gives F-term inflation with the two chiral superfields $T$ (inflaton multiplet) and $S$ (sgoldstino multiplet).
The $T$ submanifold is $SU(1,1)/U(1)$ with scalar curvature $R = -2/3$, and the no-scale structure of the K\"ahler potential is responsible for the universal expression
\begin{eqnarray}
V\ =\ M^2 \, M_{\rm Pl}^2\, \bigl(1\ -\ e^{\,-\sqrt{\frac{2}{3}}\, \varphi}\bigg)^2 \ ,
\end{eqnarray}
along the inflationary trajectory where $F_S\neq 0,~ F_T=0$, which identifies $S$ with the sgoldstino.

\subsection{Other Models}

Several examples exist with two chiral multiplets of the same sort, for which $F_S$ leads to a de Sitter  plateau  with $ F_T = 0$, while at the end of inflation $F_S = F_T = 0$ and Supersymmetry is recovered.
A class of models ($\alpha$ attractors) modify the superpotential but not the K\"ahler geometry of the original $R+R^2$ theory, which now reads
%(Kallosh, Linde, Roest)
~\cite{KLR}:
\begin{eqnarray}
W(S,T)\ =\ S\, f(T)\ , \label{sup_po}
\end{eqnarray}
with scalar curvature $R_\alpha= - \frac{2}{3\alpha}$.
Along the inflationary trajectory the potential is positive since
\begin{eqnarray}
V\ \sim \ |f|^2\ \geq \ 0 \ .
\end{eqnarray}

An alternative class of models with opposite role for K\"ahler potential and superpotential rest on the choice of eq.~(\ref{sup_po}), combined however with the trivial K\"ahler geometry corresponding to
\begin{eqnarray}
K\ =\ \frac{1}{2}\left(\Phi+\overline \Phi\right)^2\ +\ S\,\overline S\ .
\end{eqnarray}
The inflaton is now identified with $\varphi=\mathit{Im}\Phi$, thus avoiding the dangerous exponential factor $e^K$ in the supersymmetric potential. Along the inflationary trajectory
\begin{eqnarray}
V(\varphi)\ \sim \ |f(\varphi)|^2\ ,
\end{eqnarray}
so that the inflaton potential is fully encoded in the superpotential shape.

\section{Nilpotent Superfields and Sgoldstino--less Models}

In all the models reviewed so far it is difficult to exit inflation with Supersymmetry broken at a scale much lower than the de Sitter plateau (Hubble scale during inflation).
A way to solve this problem is to introduce a nilpotent (Volkov--Akulov) multiplet $S$ satisfying
~\cite{BI-R,BI-LR,BI-Casalbuoni,Seiberg} the constraint of eq.~(\ref{nilpotent}),
so that the goldstino lacks its scalar partner, which is commonly called the sgoldstino. This solves the stabilization problem and gives rise to a de Sitter plateau.

The first cosmological model with a nilpotent sgoldstino multiplet was a generalization of the Volkov-Akulov-Starobinsky supergravity~\cite{ADFS}, where
\begin{eqnarray}
W(S,T)\ =\ S\, f(T)\ , \qquad V\ =\ e^{\,K(T)}\ K^{-1}_{S\overline S} \ |f(T)|^2\ .
\end{eqnarray}
Two classes of models which incorporate separate scales of Supersymmetry breaking during and at the exit of inflation were then proposed. They rest on a trivial (flat) K\"ahler geometry
\begin{eqnarray}
K(\Phi,S)\ =\ \frac{1}{2}\Big{(}\Phi\ +\ \overline \Phi\Big{)}^2\ +\  S\,\overline S\ ,
\end{eqnarray}
but differ in their supersymmetry breaking patterns during and after inflation.
\begin{itemize}
\item In the first class of models~~\cite{Kallosh-Linde2}
\begin{eqnarray}
W(\Phi,S)\ =\ M^2\,S\,\Big{(}1\ +\ g^2(\Phi)\Big{)}\ +\ W_0\ ,
\end{eqnarray}
where $g(\Phi)$  vanishes at $\Phi=0$  and the inflaton $\varphi$ is  identified with its imaginary part. Along the inflaton trajectory  $\mathit{Re}\Phi=0$, and the potential reduces to
\begin{eqnarray}
V\ =\ M^4\ |g(\Phi)|^2\,\Big{(}2\ +\ |g(\Phi)|^2\Big{)}\ +\ V_0\ , \quad V_0\ =\ M^4\ -\ 3\, W_0^2\ .
\end{eqnarray}
Assuming   $V_0\simeq 0$, one finds
\begin{eqnarray}
m_{3/2}= \frac{H}{\sqrt{3}}\ , \ E_{SB}= |F_{S}|^{\frac{1}{2}}=\sqrt{H M_{\rm Pl}}>H\ , \
V=F_S F^S-3 W_0^2\, ,
\end{eqnarray}
while $F_\Phi=0$ during inflation $(\mathit{Re}~\Phi=0$).
\item In the second class of models
~\cite{DAZ}
% (Dall’Agata, Zwirner)
the superpotential is
\begin{eqnarray}
W(\Phi,S)\ =\ f(\Phi)\Big{(}1\ +\ \sqrt{3}\, S\Big{)} \ ,
\end{eqnarray}
which combines nilpotency and no-scale structure. Here the function $f(\Phi)$ satisfies the conditions
\begin{eqnarray}
\overline f(\Phi\ )\ =\ f(-\overline \Phi)\ , \qquad f'(0)\ =\ 0\ , \qquad f(0)\ \neq \ 0 \ .
\end{eqnarray}
The scalar potential is of no-scale type, and letting $\Phi=(a+i\varphi)/\sqrt{2}$,
\begin{eqnarray}
 F_SF^S=3 e^{a^2} |f(\Phi)|^2\ , \ V(a,\varphi)=F^\Phi F_{\Phi}=e^{\,a^2}|f'(\Phi)+a\sqrt{2}f(\Phi)|^2\, .
 \end{eqnarray}

The field $ a$ is stabilized at $a=0$,  since $f$ is an even function of $a$. During inflation $a$ gets a mass ${\cal O}(H)$ without mixing with $\Phi$ and is rapidly driven to $a=0$, so that the inflationary potential reduces to
\begin{eqnarray}
V(a=0,\varphi)\ = \ \Big{|}\,f'\left(\frac{i\varphi}{\sqrt{2}}\right)\Big{|}^2 \ , \qquad V(0,0)\ = \ 0 \ .
\end{eqnarray}
These models lack the fine-tuning of the previous class ($V_0=0$), and it
is interesting to compare the supersymmetry breaking patterns. Here $F_S$ never vanishes, and at the end of inflation
\begin{eqnarray}
F^S\,F_S\ =\ 3\, e^{G(0,0)}\ = \ 3\, m_{3/2}^2 \ .
\end{eqnarray}
In particular,
\begin{eqnarray}
\langle F^S\rangle_{\Phi=0}\ =\ \sqrt{3}\,\ \overline f(0)\ , \qquad m_{3/2} \ =\ |f(0)|\ .
\end{eqnarray}
and the inflaton potential vanishes at the end of inflation. A choice that reproduces the Starobinsky potential is
\begin{eqnarray}
f(\Phi)\ =\ \lambda \ -\ i\,\mu_1 \Phi \ +\ \mu_2\, e^{i\frac{2}{\sqrt{3}} \, \Phi}\ .
\end{eqnarray}

Interestingly, $m_a$ and $m_{3/2}$  depend on the integration constant $\lambda$, but $V$ is independent of it, and hence the same is true for $m_\varphi$.
\end{itemize}
\section{Higher-curvature Supergravity and standard Supergravity duals}

Work in this direction started with the $R+R^2$ Starobinsky model, whose supersymmetric extension was derived in the late 80’s
~\cite{Cecotti,CFPS}
%(Cecotti; Cecotti, SF, Porrati, Sabharwal)
and was recently revived in view of the new CMB data~~\cite{FKLP,FKR,FKvP,ENO}.
%(SF, Kallosh, Linde, Porrati; Farakos, Kehagias, Riotto; Kallosh, SF, Van Proeyen; Ellis, Nanopoulos, Olive, ...).
Models dual to higher-derivative theories give more restrictions than their bosonic counterparts or standard Supergravity duals.
Theories with unconstrained superfields also include the Supergravity embedding of $R^2$ duals, whose bosonic counterparts describe standard Einstein gravity coupled to a massless scalar field in de Sitter space. These theories were recently resurrected in~~\cite{KLT,AGKKLR}. The $R^2$ higher curvature Supergravity was recently obtained in both the old and new minimal formulations
~\cite{FKP2}.
%(SF, Kehagias, Porrati).
In the old-minimal formulation, the superspace Lagrangian is
\begin{eqnarray}
\alpha \,{\cal R}\, \overline {\cal R}\Big{|}_{D}\ -\ \beta \, {\cal R}^2\Big{|}_F \ ,
\end{eqnarray}
where
\begin{eqnarray}
{\cal R}\ =\ \frac{\Sigma(\overline S_0)}{S_0} \ , \qquad \overline {\cal D} _{\dot{\alpha}}{\cal R}\ =\ 0
\end{eqnarray}
is the scalar curvature multiplet, with Weyl and chiral weights $(w=1,n=1)$.  The dual standard Supergravity has K\"ahler potential and superpotential
\begin{eqnarray}
K\ =\ -\ 3 \, \log (T\ +\ \overline T\ -\ \alpha \, S\, \overline S)\ , \qquad W\ =\ T\,S\ -\ \beta \,S^3\ ,
\end{eqnarray}
and the K\"ahlerian manifold is $SU(2,1)/U(2)$. Note the rigid scale invariance of the action under
\begin{eqnarray}
T\ \to \ e^{2\, \lambda } \,T \ , \qquad S\ \to \ e^\lambda \, S \ , \qquad S_0\ \to \ e^{-\lambda} \, S_0\ .
\end{eqnarray}
If $\alpha=0$  $S$ is not dynamical, and integrating it out gives an $SU(1,1)/U(1)$ $\sigma$-model with
K\"ahler potential and superpotential
\begin{eqnarray}
   K\ =\ -\ 3 \, \log \left(T\ +\ \overline T\right)\ , \qquad W\ =\ \frac{2 \, T^{\frac{2}{3}}}{3\,\sqrt{3\,\beta}}\ .
   \end{eqnarray}
Higher-curvature supergravities can be classified by the nilpotency
properties of the chiral curvature  ${\cal R}$ . Such nilpotency constraints
give rise to dual theories with nilpotent chiral superfields~~\cite{ADFS}.
%(Antoniadis, Dudas, SF, Sagnotti).
In particular, the constraint
\begin{eqnarray}
{\cal R}^2\ =\ 0\ ,
\end{eqnarray}
in $R+R^2$
%(R is the bosonic scalar curvature)
generates a dual theory where the inflaton chiral multiplet $T$ (scalaron) is coupled to the Volkov-Akulov multiplet $S$
\begin{eqnarray}
S^2\ =\ 0\ , \qquad \overline {\cal D}_{\dot{\alpha}}\,S\ =\ 0\ .
\end{eqnarray}
For this theory (the V-A-S Supergravity), the K\"ahler potential and superpotential are
\begin{eqnarray}
K\ =\ -\ 3 \,\log\left(T\ +\ \overline T\ - \ S\,\overline S\right)\ , \qquad W\ = \ M\, S\,T\ +\ f\, S\ +\ W_0\ ,
\end{eqnarray}
respectively, and due to its no-scale structure the scalar potential is semi-positive definite
\begin{eqnarray}
 V\ =\ \frac{|M \, T\ +\ f|^2}{3\,(T\ + \ \overline T)^2}\ .  \label{V}
 \end{eqnarray}
In terms of the canonically normalized field
\begin{eqnarray}
T\ =\ e^{\sqrt{\frac{2}{3}}\,\phi}\ +\ i\,a\, \sqrt{\frac{2}{3}}\ , \qquad (\phi,a) \ \in \ \frac{SU(1,1)}{U(1)}\ ,
\end{eqnarray}
the potential eq.~(\ref{V})  becomes
\begin{eqnarray}
V\ = \ \frac{M^2}{12} \bigg(1\ -\ e^{\,-\sqrt{\frac{2}{3}}\,\phi}\bigg)^2\ +\ \frac{M^2}{18}\
e^{\,-\,2\sqrt{\frac{2}{3}}\,\phi}\, a^2 \ .
\end{eqnarray}
Here $a$ in the axion, which is much heavier than the inflaton during inflation
\begin{eqnarray}
m_\phi^2\ \simeq \ \frac{M^2}{9} \ e^{\,-\,2\,\sqrt{\frac{2}{3}}\,\phi_0}\ \ll\ m_a^2\ =\ \frac{M^2}{9}\ .
\end{eqnarray}
There are then only two natural supersymmetric models with genuine single-field  $\phi$ inflation. One is the new-minimal $R+R^2$ theory, where the inflaton has a massive vector as bosonic partner, and the V-A-S (sgoldstino-less) Supergravity just described.

Another interesting example is the sgoldstino-less version of the ${\cal R}\overline {\cal R}$  theory described before. This is obtained imposing the same
constraint  ${\cal R}^2=0$ as for the V-A-S Supergravity~~\cite{FPS}, and
is dual to the latter  with
\begin{eqnarray}
f\ =\ W_0\ =\ 0\ .
\end{eqnarray}
The corresponding potential
  \begin{eqnarray}
  V\ =\ M^2 \, \frac{|T|^2}{3\,(T\ +\ \overline T)^2}\ =\ \frac{M^2}{12}\ +\ \frac{M^2}{18}\ e^{\,-\,2\sqrt{\frac{2}{3}}\,\phi}\, a^2\ ,
  \end{eqnarray}
is positive definite and scale invariant.
This model results in a de Sitter vacuum geometry with a positive vacuum energy
\begin{eqnarray}
V(a=0)\ = \ \frac{M^2}{12}\ M_{\rm Pl}^4\ .
\end{eqnarray}

On the other hand, the Volkov-Akulov model coupled to Supergravity involves two parameters, and its vacuum energy has an arbitrary sign. The pure V-A theory coupled to Supergravity has indeed a superfield action determined by~~\cite{ADFS}
\begin{eqnarray}
K\ =\ 3\, S\,\overline S\ , \qquad W\ =\ f\,S\ + \ W_0\ , \qquad S^2\ = \ 0\ .
\end{eqnarray}
Moreover, the cosmological constant turns out to be
\begin{eqnarray}
\Lambda\ =\ \frac{1}{3}\ |f|^2\ -\ 3\, |W_0|^2 \ .
\end{eqnarray}
The full-fledged component expression of the model, including all fermionic terms, was recently worked out
~\cite{BFKvP,HY}. The higher-curvature supergravity dual~~\cite{DFKS,AM} is the standard (anti-de Sitter) supergravity Lagrangian augmented with the nilpotency constraint
\begin{eqnarray}
\left(\frac{{\cal R}}{S_0} \ - \ \lambda\right)^2\ = \ 0\ .
\end{eqnarray}
This is equivalent to adding to the action the term
\begin{eqnarray}
\sigma \left(\frac{{\cal R}}{S_0}\ - \ \lambda\right)^2 S_0^3\Big{|}_F\ ,
\end{eqnarray}
where $\sigma$ is a chiral Lagrange multiplier. A superfield Legendre transformation and the superspace identity
\begin{eqnarray}
\bigg[(\Lambda\ +\ \overline \Lambda)S_0\overline S _0\bigg]_D\ = \ \bigg[\Lambda {\cal R} S_0^2\bigg]_F\ + \ {\rm h.c}\ ,
\end{eqnarray}
 which holds up to a total derivative for any chiral superfield $\Lambda$, turn indeed the action into the V-A superspace action coupled to standard Supergravity with
\begin{eqnarray}
 f\ = \ \lambda\ -\ 3\, W_0\ .
 \end{eqnarray}
Hence, supersymmetry is broken whenever
\begin{eqnarray}
  3\, W_0\ \neq \ \lambda\ \neq\ 0 \ .
  \end{eqnarray}

In the higher-derivative formulation, the goldstino $G$ is encoded in the Rarita-Schwinger field. At the linearized level around flat space
\begin{eqnarray}
G\ =\ -\ \frac{3}{2\lambda}\left(\gamma^{\mu\nu}\,\partial_\mu \,\psi_\nu\ -\ \frac{\lambda}{2}\ \gamma^\mu\, \psi_\mu\right)\ , \qquad \delta G\ = \ \frac{\lambda}{2}\ \epsilon\ .
\end{eqnarray}
The linearized equation of motion for the gravitino reads
\begin{eqnarray}
\gamma^{\mu\nu\rho}\,\partial_\nu\,\psi_\rho\ -\ \frac{\lambda}{6}\ \gamma^{\mu\nu}\,\psi_\nu\ -\ \frac{1}{3}
\left(\gamma^{\mu\nu}\,\partial_\nu \ - \ \frac{\lambda}{2}\gamma^\mu \right)G\ =\ 0\ ,
\end{eqnarray}
and is gauge invariant under
\begin{eqnarray}
\delta\psi_\mu\ =\ \partial_\mu \epsilon\ + \ \frac{\lambda}{6}\ \gamma_\mu\, \epsilon\ .
\end{eqnarray}

Both the $\gamma$-trace and the divergence of the equation of motion yield
\begin{eqnarray}
\gamma^{\mu\nu}\, \partial_\mu\, \psi_\nu\ - \ \gamma^\mu \,\partial_\mu G\ =\ 0 \ ,
\end{eqnarray}
so that gauging away the Goldstino $G$ one recovers the standard
formulation of a massive gravitino.

Tables 1--3 summarize the various dualities linking higher-curvature supergravities in the old-minimal and new-minimal formulations with standard Supergravity.
\begin{table}
\begin{center}{\tablefont
Table 1.~ Old-Minimal Dualities\\[5pt]
$-\Phi S_0 \overline S _0\Big{|}_D+W S_0^3 \Big{|}_F, ~~~\Phi=\exp\left(-\frac{K}{3}\right)$
\\[5pt]
\begin{tabular}{ll}
\toprule
Higher Curvature Supergravity & Standard Supergravity \\
\colrule
$\Phi_H=1-h\left(\frac{{\cal R}}{S_0},\frac{\overline {\cal R}}{\overline S _0}\right)$&$\Phi_S=1+T+\overline T-h(S,\overline S)$ \\
$W_H=W\left(\frac{{\cal R}}{S_0}\right)$& $W_S=TS -W(S)$\\
\colrule
$\Phi_H=1$&$ \Phi_S=1+T+\overline T$\\
$W_H=W\left(\frac{{\cal R}}{S_0}\right)$& $ W_S=-S W'(S)+W(S)\Big{|}_{T=-W'(S)}$\\
\colrule
$\Phi_H=-\alpha \frac{{\cal R}}{S_0}\frac{\overline {\cal R}}{\overline S _0}$& $ \Phi_S=
T+\overline T-\alpha S\overline S$\\
$W_H=-\beta  \frac{{\cal R}^3}{S_0^3}$&$ W_S=TS-\beta S^3$
\\\botrule
\end{tabular}}
\end{center}
\end{table}
\vskip.3in
\begin{table}
\begin{center}{\tablefont
Table 2.~Nilpotent Old-Minimal Dualities\\[5pt]
$-\Phi S_0 \overline S _0\Big{|}_D+W S_0^3 \Big{|}_F, ~~~\Phi=\exp\left(-\frac{K}{3}\right)$
\\[5pt]
\begin{tabular}{ll}
\toprule
Higher Curvature Supergravity & Standard Supergravity \\
\colrule
$\Phi_H=1-\frac{1}{M^2} \frac{{\cal R}}{S_0}\frac{\overline {\cal R}}{\overline S _0}$&$\Phi_S=T+\overline T-S \overline S$ \\
$W_H=W_0+\xi \frac{{\cal R}}{S_0}+\sigma \frac{{\cal R}^2}{S_0^2}$& $W_S=MTS +f S+W_0$\\
& ($S^2=0, ~~f=\xi-\frac{1}{2}$)\\
\colrule
$\Phi_H=-\frac{1}{M^2} \frac{{\cal R}}{S_0}\frac{\overline {\cal R}}{\overline S _0}$&$ \Phi_S=T+\overline T-S \overline S$\\
$W_H=\sigma \frac{{\cal R}^2}{S_0^2}$& $ W_S=MTS$\\
 & ($S^2=0$)\\
\colrule
$\Phi_H=1$& $ \Phi_S= 1- S\overline S$\\
$W_H=W_0+\sigma \left(\frac{{\cal R}}{S_0}-\lambda\right)^2$&$ W_S=fS+W_0$\\
& ($ S^2=0,~f=\lambda -3 W_0$)
\\\botrule
\end{tabular}}
\end{center}
\end{table}
\vskip.3in
\begin{table}
\begin{center}{\tablefont
Table 3.~New-Minimal Dualities\\[5pt]
$\Phi=\exp\left(-\frac{K}{3}\right)$
\\[5pt]
\begin{tabular}{ll}
\toprule
Higher Curvature Supergravity & Standard Supergravity \\
\colrule
$L\log \left(\frac{L}{S_0\overline S _0}\right)\Big{|}_D $&$\Phi_S=-U \exp U$ \\
$W_\alpha \left(\frac{L}{S_0\overline S _0}\right) W^\alpha \left(\frac{L}{S_0\overline S _0}\right) \Big{|}_F$ & $ W_\alpha(U)W^\alpha(U)$\\
\colrule
 & $\Phi_S=(T+\overline T) \exp V $\\
$W_\alpha \left(\frac{L}{S_0\overline S _0}\right) W^\alpha \left(\frac{L}{S_0\overline S _0}\right) \Big{|}_F$ & $ W_\alpha(V)W^\alpha(V)$
\\\botrule
\end{tabular}}
\end{center}
\end{table}

\section{Orthogonal Nilpotent Superfields}

We have seen so far that simple models of inflation, and in particular the supersymmetric version of the Starobinsky model, rest on a pair of chiral multiplet, the sgoldstino multiplet $S$ and the inflaton multiplet $T$. Sgoldstino-less models are obtained by replacing $S$ by a nilpotent superfield ($S_{NL}^2=0$), which is the local version of the V-A multiplet. This setting should correspond to a linear model where the scalar partners of the goldstino are infinitely heavy, so that the sgoldstino becomes a non-dynamical
composite field. Following~~\cite{BFZ,Seiberg},
other types of constraints can be imposed, which remove
other degrees of freedom from the $T$ multiplet. The most interesting of
them is the orthogonality constraint
~\cite{FKT,CKLinde,DAF}
\begin{eqnarray}
S_{NL}\,(T_{ONL}\ - \ \overline T _{ONL})\ =\ 0\ , \qquad (S_{NL}^2\ =\ 0) \ , \label{ortho_con}
\end{eqnarray}
which also implies
\begin{eqnarray}
\left(T_{ONL} \ - \ \overline T _{ONL}\right)^3\ = \ 0 \ .
\end{eqnarray}
This constraint removes the inflatino (spin-$1/2$ partner of the inflaton), as well as the sinflaton (spin-$0$ partner of the inflaton), so that this description should correspond to a regime where the inflatino and the sinflaton are both infinitely heavy.

The new aspect of these ``non-chiral” orthogonality constraints"  is that the $T$-auxiliary field $F_T$ becomes nilpotent, and therefore fails to contribute to the scalar potential, which takes the form
\begin{eqnarray}
V(\varphi\, =\, \mathit{Re}~T)\ =\ f^2(\varphi)\ -\ 3\, g^2(\varphi)\ , \label{VONL}
\end{eqnarray}
for a quadratic K\"ahler potential and a superpotential of the form
\begin{eqnarray}
W(S_{NL},T_{ONL})\ =\ S_{NL}\,f(T_{ONL})\ +\ g(T_{ONL})\ .
\end{eqnarray}

The potential $V$ in eq.~(\ref{VONL})  may or may not reproduce the inflaton trajectory for models with a ``linear” $T$ multiplet".
This setting presents an advantage with respect to the linear $T$ model, because it eliminates the sinflaton, thus bypassing the problems related to its stabilization. It also avoids goldstino-inflatino mixing, which makes matter creation in the Early Universe very complicated. In the unitary gauge, the inflatino field simply vanishes, since it is proportional to the goldstino~~\cite{FKT,DAFZ}.

In Table 4 we collect the various orthogonality constraints.
%\vskip.3in
%
\begin{table}
\begin{center}
{\tablefont Table 4.~Orthogonality constraints with $S_{NL} ~(S_{NL}^2=0$)
\begin{tabular}{ll}
\toprule
 $S_{NL}(T_{ONL}-\overline T _{ONL})=0$& sgoldstino-less, inflatino-less, sinflaton-less\\
(implies $(T_{ONL}-\overline T _{ONL})^3=0$)& \\ \\
$S_{NL}\overline T'_{ONL}=\mbox{chiral}$  & sgoldstino-less, inflatino-less\\
(implies $S_{NL}\overline {\cal D}_{\dot{a}} T'_{ONL}=0)$&\\ \\
  $S_{NL} T''_{ONL}=0$& sgoldstino-less, scalar-less\\
  (implies $(T''_{ONL})^3=0$)& \\ \\
    $S_{NL} W_\alpha(V_{ONL}=0$&sgoldstino-less, gaugino-less\\
%\colrule \\
\botrule
\end{tabular}}\label{t1}
\end{center}
\end{table}
The supergravity model for a matter multiplet $T$ corresponding to the constraint
$ST=0$ was derived in \cite{DAFZ}. This model has been recently shown \cite{wrase_16} to describe the effective dynamics of a fermion, other than the ${\cal N}=1$ goldstino, which lives on a
$\overline{D}3$-brane world volume.

\section{Conclusions and Outlook}
The orthogonality constraints in eq.~(\ref{ortho_con}) and the resulting scalar potential in eq.~(\ref{VONL}) allow the  construction of  MSIM (minimal supersymmetric inflationary models), which accommodate, with appropriate fine tuning, dark energy (cosmological constant $\Lambda$),
the supersymmetry breaking scale $m_{3/2}$, and the inflationary Hubble scale $H$~~\cite{CKLinde}.
A simplified class of models is obtained with (in $M_{\rm Pl}$ units)
\begin{eqnarray}
g(\varphi)\ =\ g_0\ =\ m_{3/2}\ , \qquad f(\varphi)\ =\ H\, f_I(\varphi)\ +\ f_0\ ,
\end{eqnarray}
where $\varphi$ is the appropriate canonically normalized scalar field, whenever the K\"ahler potential is not quadratic but has the more general form as in refs~~\cite{FKT,CKLinde}. Here, $f_I(\varphi)$ is a function with the property $f_I(\varphi)\to 1$, (for $\varphi$ large) while at the extremum of the
potential $\varphi=0$, $f_I(\varphi)=0$. Hence, the scalar potential
satisfies $V(\varphi=0)=f_0^2-3m_{3/2}^2=\Lambda$, while for large $\varphi$, $V(\varphi)\to H^2$ (as $\varphi\to \infty$), for values of the parameters such that $\Lambda\approx 10^{-120}$, $m_{3/2}\approx 10^{-16}$ and  $H=10^{-5}$, where we took the SUSY breaking scale at the end of inflation (approximate Minkowski spacetime) to be at the {\rm TeV} scale as a minimal value, which is inspired by the current LHC results.

Finally, we would like to note that microscopic models which may yield in suitable limits the non-linear realisations considered so far have been proposed in refs~~\cite{KKMM,FKvPW,DDF} and
%coupling to matter of sgoldstino-less models was considered in ref~~\cite{KLW}.
matter couplings to the inflation sector, with and without non-linear superfields, were considered in  \cite{Kallosh-Linde2,DAZ,DFKS,FP-2,KLW}.

This review reflects the lecture presented by SF at the 2016 Memorial  Meeting for Prof.~Salam, and overlaps in part with \cite{other_revs}.
\vskip.2in
\noindent
{\bf Acknowledgment}: We are grateful to I.~Antoniadis, G.~Dall'Agata, E.~Dudas, F.~Farakos, P.~Fr\'e, R.~Kallosh, A.~Linde, M.~Porrati, A.~Riotto, A. Sorin,  J.~Thaler, A.~Van Proeyen,  T. Wrase and F.~Zwirner for useful discussions and collaboration on related projects. SF is supported in part by INFN-CSN4-GSS, while AS is supported in part by Scuola Normale Superiore and by INFN-CSN4-STEFI.

\appendix{\bf Constrained Superfields and ${\cal N}=1$ Supersymmetry}

Non-linear constraints involving a pair of chiral superfields $(X,W_\alpha)$, $(X,U_{\dot{a}})$ can have solutions that differ sharply from the V-A case. Here $W_\alpha,\ U_{\dot{a}}$ are the chiral superfields
\begin{eqnarray}
&&W_\alpha\ = \-\ \frac{1}{4}\ \bar D^2 D_\alpha V \ , \quad \mbox{gauge field-strength multiplet}\ , \label{W}\\
&& U_{\dot a}\ = \ \bar D _{\dot a} L \ , \quad \mbox{$L$ linear (or tensor) multiplet, $D^2 L=\bar D^2 L=0$}\ .\label{L}
\end{eqnarray}
The (chiral) constraints in question are
\begin{eqnarray}
1) &&X^2\ =\ 0\ , \quad X W_\alpha=0\ , \label{XW} \\
2) &&X^2\ =\ 0\ , \quad X U_{\dot a}=0 \ , \quad \mbox{(or $XL=$ chiral)}\ . \label{XU}
\end{eqnarray}
 ${\cal N}=1$ supersymmetry is broken solving $X^2=0$, with the
V-A solution
\begin{eqnarray}
 X=X_{NL}\ =\ S\ =\ \left(\frac{G^2}{2F},G_\alpha,F\right)\  ,
 \end{eqnarray}
and then solving the second portions of eqs~(\ref{XW},\ref{XU}), 
\begin{eqnarray}
 X_{NL}W_\alpha\ =\ 0\ , \quad  X_{NL} U_{\dot a}=0\ . \label{XWU}
 \end{eqnarray}
Using the fact that $W_\alpha,\ U_{\dot a}$ have components
\begin{eqnarray}
&&W_\alpha\ =\ \left(\lambda _\alpha, ~L_{\alpha}^{\beta}\ =\ \delta_{\alpha}^{\beta}D-\frac{i}{2}(\sigma^\mu\bar \sigma ^\nu)^\beta_\alpha F_{\mu\nu},
%~F_{\alpha\beta}+i\epsilon_{\alpha\beta}D,
~\partial_{\alpha\dot{a}}\bar \lambda ^{\dot a}\right)\  , \\
&& U_{\dot a}\ =\ \left( \bar{\chi}_{\dot a}, ~ \Lambda _{{\dot a}\beta}=\sigma^\mu _{\dot{a}\beta}
(H_\mu+i \partial_\mu \phi),~\partial_{\alpha\dot{a}}\bar \chi ^{a}\right)\  ,
\end{eqnarray}
where
\begin{eqnarray}
F_{\mu\nu}\ =\ \partial_\mu A_\nu-\partial_\nu A_\mu\ , ~~~H_\mu\ =\ \frac{1}{3!}\ \epsilon_{\mu\nu\rho\sigma}\,\partial^\nu\, b^{\rho\sigma}\ ,
\end{eqnarray}
the constraints leave free the bosonic fields but express the \emph{gaugino} (and \emph{tensorino})
in terms of   the V-A $G$ goldstino ~\cite{Seiberg} according to
\begin{eqnarray}
&&\lambda_\alpha \ = \ i\, L_{\alpha \beta} \ \frac{G^\beta}{\sqrt{2}F}\ +\ {\cal O}(G^2)\ , \\
&&\bar{\chi}_{\dot{a}}\ =\ -\,i \, \Lambda_{\dot{\alpha} \beta} \frac{G^\beta}{\sqrt{2}F}\ +\ {\cal O}(G^2)\ .
\end{eqnarray}

The full solution can be obtained from the last component of the constraints by iteration. The constraint on the linear multiplet was considered in ~\cite{DDF}, and has the effect of leaving $(\phi,~b_{\mu\nu})$ in the spectrum. However, there is another solution to the constraints, where instead the chiral multiplet $X$ is not the V-A multiplet but the constraints in eqs.~(\ref{XW},\ref{XU}) can be used to express $X$ in terms of $W_\alpha$ (or $U_{\dot{a}})$. This is the case of the supersymmetric Born--Infeld
 and the non--linear tensor multiplet constraints of Bagger and Galperin~\cite{other_revs,FSY,BG1,BG2} where
\begin{eqnarray}
1) && X\ =\ \frac{W^\alpha \,W_\alpha}{ m\ -\ \bar D^2 \bar X}\ , \\
2)&& X\ =\ \frac{U^{\dot \alpha}\, U_{\dot \alpha}}{ m\ -\ \bar D^2 \bar X}\ .
\end{eqnarray}
The resulting Lagrangians, which are simply the F-components of $X$, describe a non--linear theory with ${\cal N}=2$ spontaneously broken to ${\cal N}=1$.

\end{document}